\documentclass[conference]{IEEEtran}
\IEEEoverridecommandlockouts
\usepackage{cite}
\usepackage{amsmath,amssymb,amsfonts}
\usepackage[figurename=Fig.]{caption}
\usepackage[tablename=Tab.]{caption}
\usepackage{algorithmic}
\usepackage{graphicx}
\usepackage{tabularx}
\usepackage{booktabs}
\usepackage{subcaption}
\usepackage{textcomp}
\usepackage{xcolor}
\usepackage{url}
\def\BibTeX{{\rm B\kern-.05em{\sc i\kern-.025em b}\kern-.08em
T\kern-.1667em\lower.7ex\hbox{E}\kern-.125emX}}

\makeatletter

\def\ps@IEEEtitlepagestyle{%
  \def\@oddfoot{\mycopyrightnotice}%
  \def\@evenfoot{}%
}
\def\mycopyrightnotice{%
  {\footnotesize 979-8-3503-9114-5/24/\$31.00 \textcopyright2024 IEEE\hfill} %
  \gdef\mycopyrightnotice{}
}

\begin{document}

\title{Multi-Reference  Generative Face Video Compression with Contrastive Learning}

\author{\IEEEauthorblockN{{Goluck Konuko}
\IEEEauthorblockA{\textit{Laboratoire de Signaux et Syst\`emes (L2S)} \\
\textit{CenraleSupelec - Universit\'e Paris-Saclay}\\
Paris, Frame \\
goluck.konuko@centralesupelec.fr}}
\and
\IEEEauthorblockN{Giuseppe Valenzise}
\IEEEauthorblockA{\textit{Laboratoire de Signaux et Syst\`emes (L2S)} \\
\textit{CNRS}\\
Paris, France \\
giuseppe.valenzise@centralesupelec.fr}}

\maketitle

\begin{abstract}
Generative face video coding (GFVC) has been demonstrated as a potential approach to low-latency, low bitrate video conferencing. GFVC frameworks achieve an extreme gain in coding efficiency with over 70\% bitrate savings when compared to conventional codecs at bitrates below 10kbps. In recent MPEG/JVET standardization efforts, all the information required to reconstruct video sequences using GFVC frameworks are adopted as part of the supplemental enhancement information (SEI) in existing compression pipelines. In light of this development, we aim to address a challenge that has been weakly addressed in prior GFVC frameworks, i.e., reconstruction drift as the distance between the reference and target frames increases. This challenge creates the need to update the reference buffer more frequently by transmitting more Intra-refresh frames, which are the most expensive element of the GFVC bitstream. To overcome this problem, we propose instead multiple reference animation as a robust approach to minimizing reconstruction drift, especially when used in a bi-directional prediction mode. Further, we propose a contrastive learning formulation for multi-reference animation. We observe that using a contrastive learning framework enhances the representation capabilities of the animation generator. The resulting framework, MRDAC (\textbf{M}ulti-\textbf{R}eference \textbf{D}eep \textbf{A}nimation \textbf{C}odec) can therefore be used to compress longer sequences with fewer reference frames or achieve a significant gain in reconstruction accuracy at comparable bitrates to previous frameworks. Quantitative and qualitative results show significant coding and reconstruction quality gains compared to previous GFVC methods, and more accurate animation quality in presence of large pose and facial expression changes. The source code will be available at \textcolor{teal}{\url{https://github.com/Goluck-Konuko/animation-based-codecs}}
\end{abstract}


\section{Introduction}
Generative Face Video Coding (GFVC) has emerged as a promising area of research for low-bitrate transmission and storage of talking-head video sequences. These sequences are ubiquitous in video conferencing, but also play a significant role in the metaverse, i.e., for educational, entertainment and social media video applications.  
GFVC coding frameworks~\cite{konuko2020dac,wang2021fv2v,chen2022cfte,konuko2022hdac,konuko2023rdac} have been shown to provide significant coding efficiency gains for low-bitrate video conferencing. 
Initial GFVC methods predict frames in a video based on a single reference frame, typically the first of the sequence or of the group of pictures (GOP)~\cite{konuko2020dac}. 
A significant challenge in GFVC is maintaining prediction accuracy as the temporal distance between the reference frame and target frames increases. This is particularly problematic in scenes with large pose variations. Previous methods have explored various strategies to address this, such as improving sparse motion representation vectors~\cite{chen2022cfte}, multiple reference coding with feature fusion~\cite{volokitin2022mvac,wang2022drcfte}, and side information enhancement~\cite{konuko2022hdac}, and residual coding~\cite{konuko2023rdac}. 
Despite these efforts, the issue of prediction quality deterioration as target frames move further in time from the references remains unsolved. 

In this paper, we focus on advancing the representation capability of GFVC frameworks. We consider the use of \textit{multiple references} to refine prediction accuracy.
Specifically, we propose a simple yet powerful reformulation to the optimization objective for the generative animation architecture to increase its accuracy in motion prediction and, consequently, its performance under diverse coding conditions. To this end, we present a contrastive learning~\cite{chen2020cl} framework for multi-reference animation-based talking-head reconstruction. We name the proposed method \textbf{MRDAC}, for Multi-Reference Deep Animation Codec. Note that while GFVC frameworks are typically designed for low-delay video processing using a single reference, multi-reference GFVC can be used both in the low-delay mode as well as in a bi-directional prediction mode.

By treating feature representations obtained from multiple references as augmentations of each other and enforcing a similarity constraint, we significantly enhance motion prediction accuracy and overall video quality. This differs from the previous frameworks that use multiple reference frames to independently predict the target frame and then either combine their predictions in the feature space of the generator~\cite{volokitin2022mvac} or in the pixel space after the generation process~\cite{wang2022drcfte}.
Our approach, by directly minimizing the distance between the feature representations obtained from multiple reference frames, ensures that each prediction has a higher accuracy, leading to globally better predictions compared to previous art. 
The experimental results confirm the superiority of the proposed approach in terms of perceptual and pixel fidelity scores compared to other GFVC frameworks proposed in the literature. In summary, the main contribution of this paper to the topic of GFVC: A contrastive learning formulation for multi-reference animation with a goal of improving the representation efficiency.

\begin{figure*}
    \centering
\includegraphics[width=0.99\linewidth]{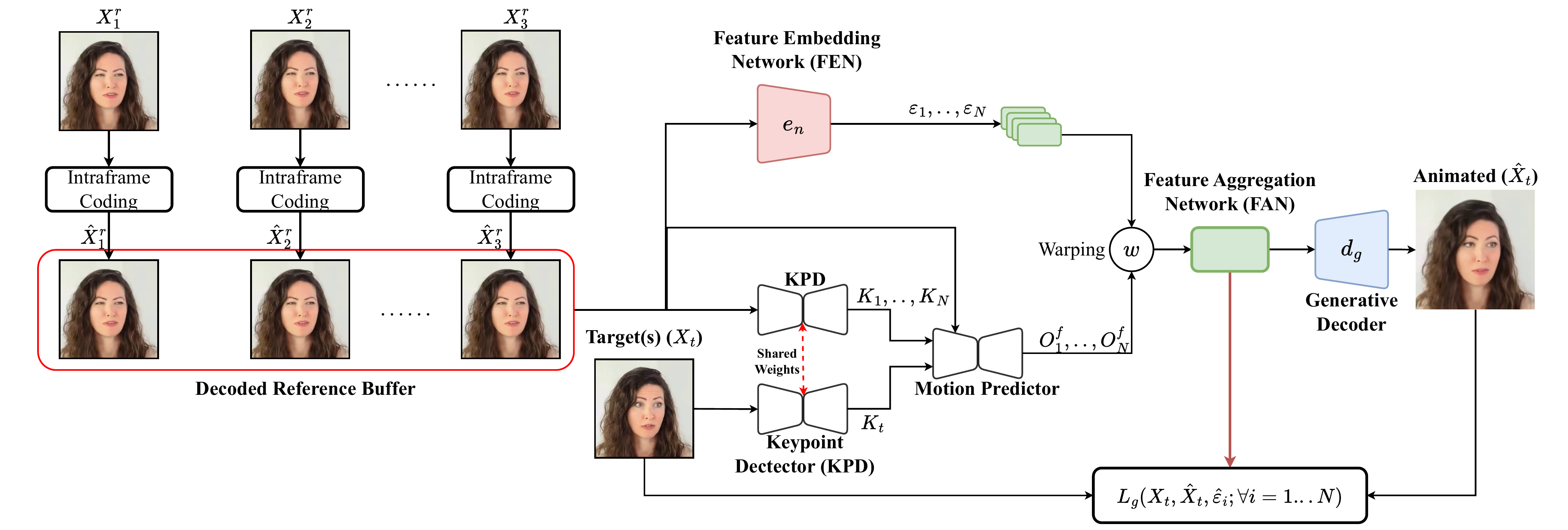}
\caption{\textbf{Proposed multi-reference deep animation codec (MRDAC)}: For a target video sequence, a group of keyframes ($X^{r}_{1},\ldots,X^{r}_{N}$) are used to predict an aggregate feature representation from which a target frame is predicted. The motion information between each reference frame and the target is predicted independently. We propose a loss function that maximizes agreement between the feature representation $\varepsilon_i$ derived from each reference frame. }
\label{fig:pipeline}
\end{figure*}

\section{Related Work}
\subsection{Generative Face Video Compression}
Animation-based representations for talking-head videos, originally proposed for avatar animation and speaker self-reenactment~\cite{siarohin2019fom}, have recently been adapted for video conferencing compression~\cite{konuko2020dac,wang2021fv2v,chen2022cfte,konuko2022hdac,konuko2023rdac,volokitin2022mvac,wang2022drcfte}. Initial methods such our previously proposed DAC~\cite{konuko2020dac} and similar works~\cite{wang2021fv2v,chen2022cfte} used a single reference frame along with motion landmarks from subsequent frames to reconstruct the video sequence, achieving competitive coding performance at ultra-low bitrates. To overcome the loss of temporal coherence in the animation, in~\cite{konuko2022hdac} we proposed a hybrid, layered coding scheme consisting of a low-quality conventional HEVC stream and an animation-based stream similar to DAC. A variant of this solution has been recently submitted to GFVC standardization in JVET-AH0114~\cite{JVETAH0114}. Animation can also be interpreted as a spatial predictor, with residual coding added to mimic classical closed-loop video codecs~\cite{konuko2023rdac}. These codecs~\cite{konuko2022hdac,konuko2023rdac} can, in principle, reduce or eliminate temporal drift in animation. However, the extra transmission cost of the low-quality HEVC or residual bitstream limits their functionality at extremely low bitrates. Subsequent studies, such as~\cite{chen2022cfte}, introduced efficient bitstream representations for motion keypoints, while~\cite{chen2022som} developed a temporal learning strategy to improve the temporal consistency of animation-based reconstructions. However, these methods struggle with occlusions and dis-occlusions between the reference and target frames and are prone to drift as the temporal distance from the reference frame increases due to their open-loop nature.
The use of multiple reference frames in animation-based reconstruction has been previously explored in~\cite{volokitin2022mvac, wang2022drcfte}. These studies clearly demonstrate that employing more than one reference frame can enhance prediction accuracy compared to single-reference GFVC frameworks. However, these approaches independently optimize each reference frame to predict the target frame and then fuse the predictions either in the feature space~\cite{volokitin2022mvac} or in the pixel space~\cite{wang2022drcfte}. The closest method to our proposed one is the MVAC framework~\cite{volokitin2022mvac}, which however requires an initial search for suitable reference frames throughout the video sequence, with a consequent potential coding delay. Our approach instead samples previously decoded reference frames, and incorporates a contrastive learning objective on the feature representations derived from multiple reference frames. We implement a minimal delay bi-prediction mechanism by buffering a small number of frames.

\subsection{Contrastive Learning}
Contrastive learning~\cite{chen2020cl,khac2020cl-review} is a widely recognized framework for representation learning, notable for its effectiveness in self-supervised learning tasks. This framework centers on distinguishing between similar and dissimilar pairs of data points. Key components include the formation of positive and negative pairs, a similarity measure, and a contrastive loss function. Positive pairs consist of different augmented views of the same data instance, while negative pairs comprise augmented views from different instances~\cite{khac2020cl-review}. Using a similarity measure, such as cosine similarity, the model is trained to bring positive pairs closer in the representation space while pushing negative pairs apart.

In the context of multi-reference animation for face video compression, we treat feature representations from two or more reference frames, which are transformed to predict a common target frame, as positive pairs. All other samples in the batch are considered negative pairs. Our hypothesis is that by minimizing the distance between positive pairs, we maximize the ``agreement'' between multiple references on the optimal representation for generating a given target frame. This approach enhances the robustness and accuracy of the motion prediction, leading to higher quality reconstructions. 

\section{Proposed Method}
We present our multi-reference animation framework, illustrated in Fig.~\ref{fig:pipeline}, as follows: we first give a summary of multi-reference animation with a generative autoencoder in Section~\ref{subsec: mr-animation}, highlighting the benefits of using multiple references for improved talking-head reconstruction. Subsequently, we explain our contrastive learning strategy for generative animation in Section~\ref{subsec:contrastive}, where we derive a contrastive learning formulation for multi-reference animation.

\subsection{Multi-Reference Face Animation}
\label{subsec: mr-animation}
We consider the task of reconstructing a target frame $X_t$ from a video sequence with frames $X_{0 \to T}$. Existing animation-based reconstruction frameworks use a single reference frame $X^{r}_{0}$ to approximate the target frame $X_t$. However, this approach often yields limited accuracy in the presence of large head motions and occlusions. Utilizing multiple reference frames ($X^{r}_{0,\ldots,N}$) sampled from the video sequence ($X_{0 \to T}$) provides a more robust structural prior, improving reconstruction accuracy under significant motion conditions. Our method builds on the autoencoder-based approach proposed in the first-order model~\cite{siarohin2019fom}. In this approach, sparse motion landmarks are extracted from each of the reference frames ($K^{r}_{0,\ldots,N}$) and the target frame ($K_t$). To reconstruct the target frame $X_t$, the animation network takes as input the reference frames and the corresponding motion landmarks.

The animation generation process begins with predicting a dense optical flow ($O^f_{r \to t}$) and an occlusion mask ($O^m_{r \to t}$)
between each reference frame and the target frame through a dense motion generator (DMG). Next, the predicted optical flow maps are used to warp the features of the input reference frames independently, and the corresponding occlusion masks are applied to the deformed reference features. This transformation is represented as follows:
\begin{equation}
\hat{\varepsilon}_r = 
O^{m}_{r \to t} 
\cdot 
\left( O^{f}_{r \to t} \circ \varepsilon_r \right), 
\quad \forall r \in 
\{0,1,\ldots,N\},
\label{eq:epsilon}
\end{equation}
where $\hat{\varepsilon}_r$ is the deformed feature for reference $r$, \(\circ\) denotes deformation through a grid sampling operation and $\cdot$ is a tensor dot product.
The deformed features of the reference frames, $\hat{\varepsilon}_0, \hat{\varepsilon}_1, \ldots, \hat{\varepsilon}_N$, serve as approximations of the feature representation of the target frame. These need to be aggregated into a single representation to predict an approximation of the target frame \(\hat{X}_t\). While previous work used a max pooling approach due to its permutation invariance, we propose an enhancement by incorporating a weight vector in the aggregation process. 
During training, this weight vector is constructed based on the relative temporal distance between each reference frame and the current target frame. The rationale is that a reference frame temporally closer to the target frame is generally closer in pose and thus provides a better prediction. Therefore, before applying max pooling, we scale the reference feature activations 
$\hat{\varepsilon}_0, \hat{\varepsilon}_1, \ldots, \hat{\varepsilon}_N$ 
by the aggregation weights 
$\lambda_0, \lambda_1, \ldots, \lambda_N$:
\begin{equation}
\varepsilon^{*} = \max(\lambda_0  \hat{\varepsilon}_0, \lambda_1  \hat{\varepsilon}_1, \ldots, \lambda_N  \hat{\varepsilon}_N)
\end{equation}
where $\max$ denotes feature-wise max pooling.
This weighted aggregation is more robust than simple maxpooling, as it prioritizes reference frames closer to the target. Note that the temporal distance is used to manually derive the weight vector for each batch sample during training and is not a learnable feature.

\subsection{Contrastive Learning for Multi-Reference Animation}
\label{subsec:contrastive}
Contrastive learning is a self-supervised technique that learns useful representations by contrasting positive pairs against negative pairs. The core idea is to bring similar (positive) pairs closer in the representation space while pushing dissimilar (negative) pairs farther apart~\cite{khac2020cl-review}. For multi-reference animation, we treat the deformed feature representations from pairs of reference frames, such as \((\hat{\varepsilon}_0, \hat{\varepsilon}_1)\), as positive pairs. Each reference frame is used to compute a feature representation that independently predicts the target frame \(X_t\). By treating \(\hat{\varepsilon}_0\) and \(\hat{\varepsilon}_1\) as augmentations of each other and computing a contrastive loss between them, we enforce a degree of ``agreement'' between the reference frames when predicting a common target frame.

The contrastive loss function for a sample $i$ in a batch is computed as follows:
\begin{equation}
  \mathcal{L}_{i} = -\log \frac{\exp(\text{sim}(\varepsilon^{i}_1, \varepsilon^{i}_2)/\tau)}{\sum_{j=1}^{B} \exp(\text{sim}(\varepsilon^{i}_1, \varepsilon^{j}_2)/\tau)} 
\label{eq:loss}
\end{equation}

where \(\text{sim}(a, b)\) is a pairwise distance measure, \(\tau\) is a temperature parameter, and \(B\) is the number of samples in the batch. This contrastive loss, combined with perceptual and equivariance losses~\cite{siarohin2019fom}, ensures lower variance in predictions from multiple references.

This approach is particularly effective even when reference frames exhibit large pose differences both between themselves and with the target frame. By enforcing similarity in their deformation towards the target, we enhance the robustness of the motion generator in two key ways. First, the motion generator learns an increased motion range and complexity which increases its overall capability to predict dense motion vectors even under conditions with large variations between the reference frame and target frames. We observe this in the increased accuracy under single reference animation mode using the proposed framework.
Second, the in-painting mechanism is improved  using multiple references since the motion generator learns to produce better occlusion masks for each reference frame. This is observed in higher pixel accuracy for both foreground and background details.
This dual enhancement leads to more accurate and consistent predictions, improving the quality of the generated frames even under challenging conditions.

\section{Results and Discussion}

\begin{figure*}[t]
\centering
\begin{subfigure}{.48\linewidth} 
\centering
\includegraphics[width=.99\linewidth]{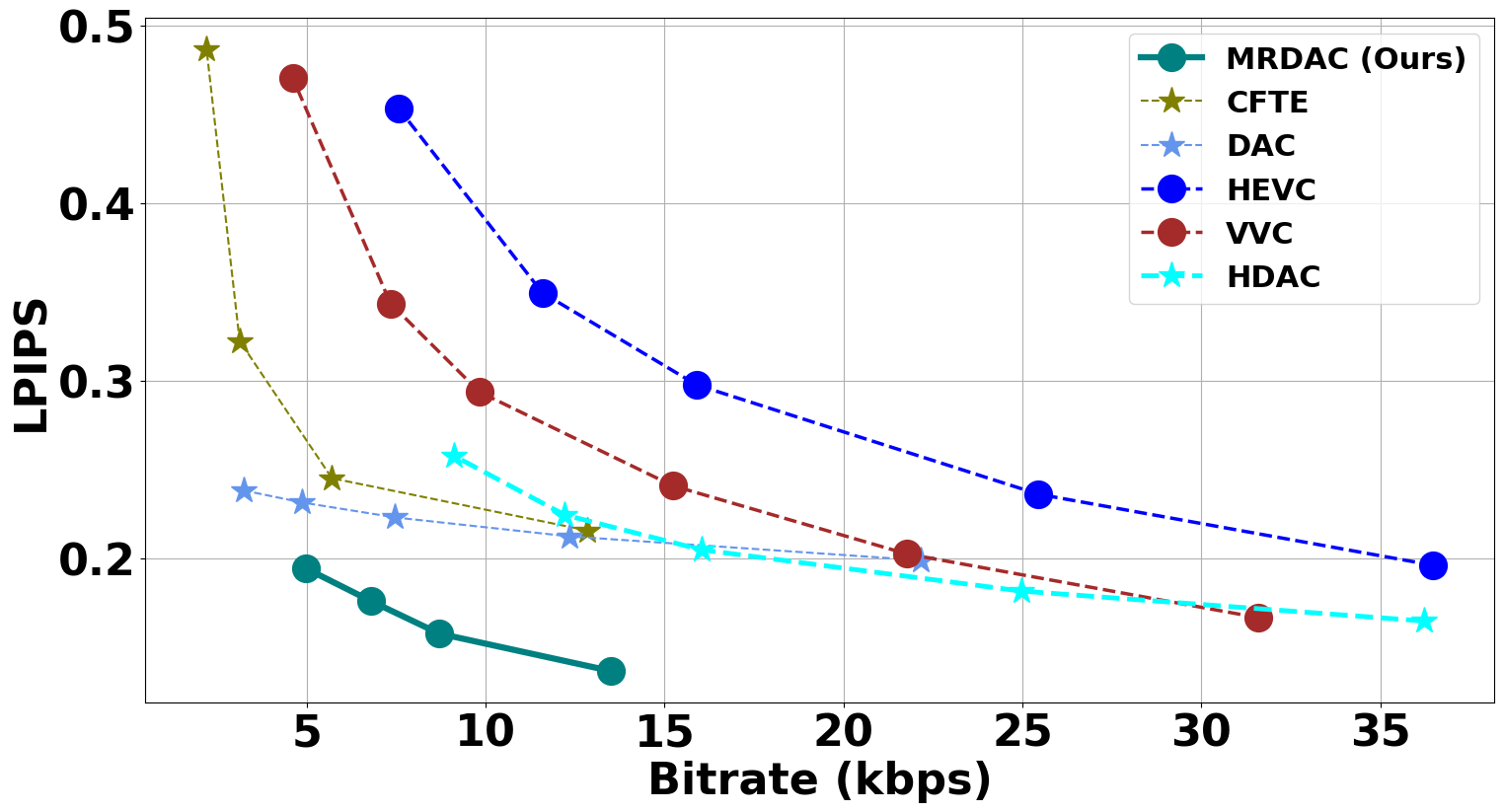}
    \label{fig:lpips-rd}
\end{subfigure}
\begin{subfigure}{.48\linewidth} 
    \centering
\includegraphics[width=.99\linewidth]{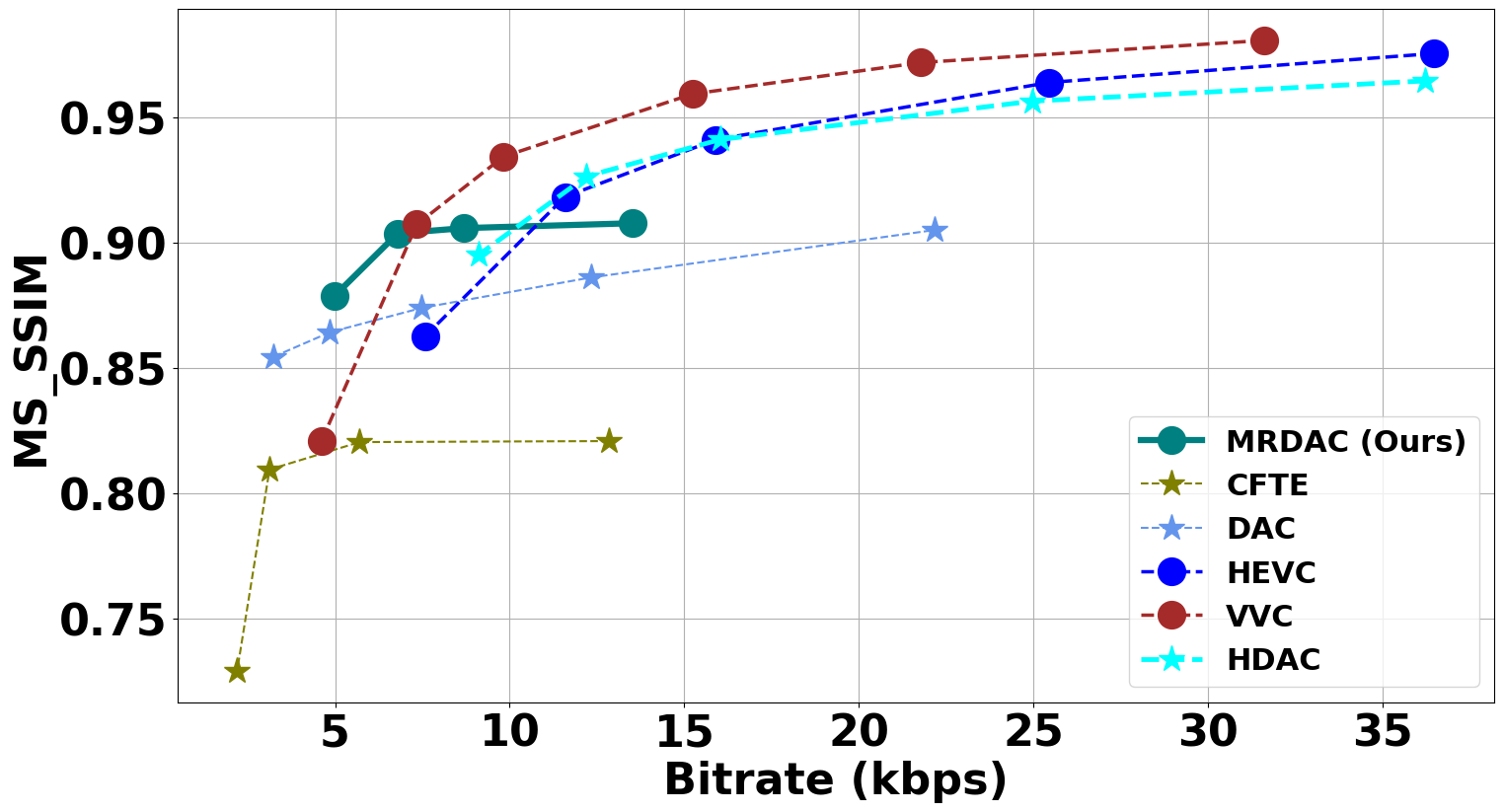} 
    \label{fig:ms-ssim}
\end{subfigure}

\caption{\textbf{Average RD Performance}: Within the standardized GFVC setting, a long sequence is animated from a single reference frame leading to ultra-low bitrate compression. Our framework progressively introduces new reference frames and jointly uses them to animate the subsequent frames leading to a significantly higher average reconstruction performance.}
\label{fig:rate-distortion}
\end{figure*}

\subsection{Datasets and Model Training}
The model architecture is similar to our prior work DAC~\cite{konuko2020dac} with the addition of the feature aggregation module proposed by~\cite{volokitin2022mvac}. At training time, we sample four random frames as reference frames and one target frame. We compute a contrastive loss between features derived from any two reference frames, in addition to employing perceptual losses, equivariance loss, and GAN losses common in other animation frameworks. The training dataset comprises 18k talking-head videos from the VoxCeleb2 dataset. The model is trained with 5 random samples per video, resulting in 90k samples per epoch over 50 epochs. The AdamW optimizer, with a learning rate of 2e-4, is utilized with a batch size of 48. Data augmentations are applied as described in~\cite{siarohin2019fom}. The model training is done on 4 NVIDIA A100 with 32GB memory.


\subsection{Evaluation Protocol}
Our evaluation protocol assesses the performance of our proposed framework against GFVC methods using test videos that exhibit a wide range of poses, expressions, and talking head motion patterns. We utilize perceptual quality metrics to measure reconstruction accuracy at various levels, from pixel fidelity to higher-level features. Specifically, we employ FSIM and MS-SSIM\footnote{\url{https://gitlab.com/wg1/jpeg-ai/jpeg-ai-qaf}} as low-level, pixel fidelity metrics. Additionally, we include learning-based metrics such as LPIPS~\cite{zhang2018lpips} and MS-VGG~\cite{simonyang2015vgg}, calculated on a frame-by-frame basis, as well as the DISTS~\cite{ding2022dists} metric, which evaluates texture and style preservation in the reconstructed videos. Furthermore, we incorporate the VMAF\footnote{\url{https://github.com/Netflix/vmaf}} metric, which includes Detail Loss Metric (DLM) to assess loss of detail in textures and motion-based metrics to consider the effect of motion on perceived quality. The motion-based metric is particularly important for our work, as our animation models reconstruct talking head sequences where motion plays a crucial role in perceived visual quality and realism. These comprehensive metrics ensure a thorough assessment of the framework's performance across different abstraction levels.

\begin{figure*}[t]
 \def\im#1{ \includegraphics[width=25mm,height=25mm]{#1}}
     \centering
   \setlength\tabcolsep{0.5 pt}
   \renewcommand{\arraystretch}{0.2}
     \begin{tabular}{*{7}{c}}
    G. TRUTH  & VVC & HDAC & CFTE & DAC & MVAC & \textcolor{blue}{MRDAC (Ours)} \\
    \im{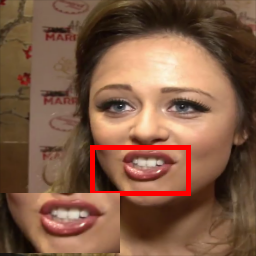} &   
    \im{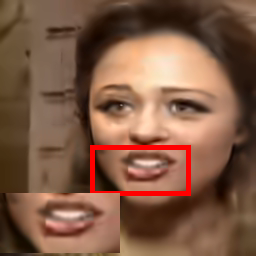} &  
    \im{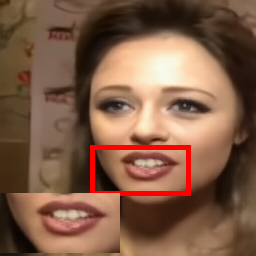} &  
    \im{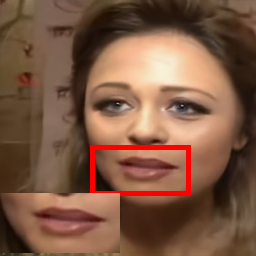} &  
    \im{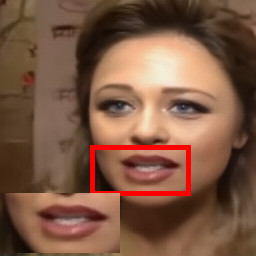} & 
    \im{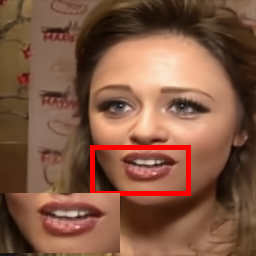} & 
    \im{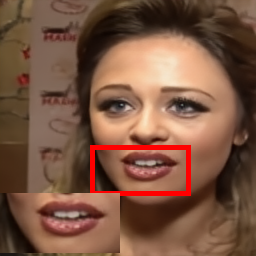}  
    \\
    \im{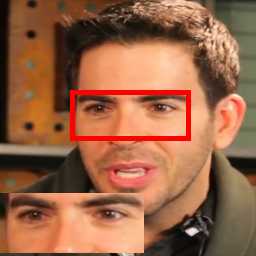} &
    \im{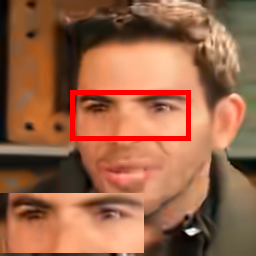} &
    \im{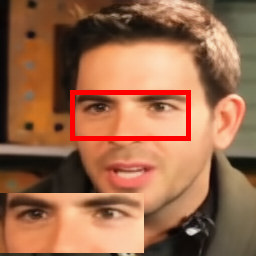} &
    \im{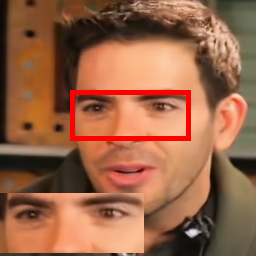} &
    \im{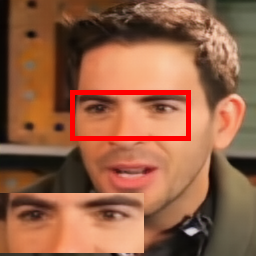} &
    \im{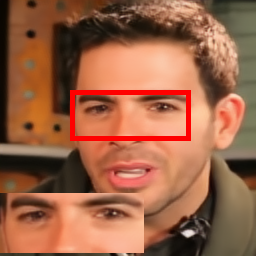} &
    \im{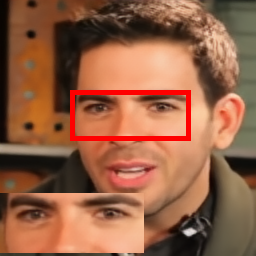}
    \\
    \im{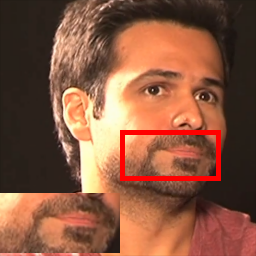} &
    \im{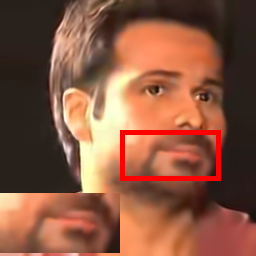} &
    \im{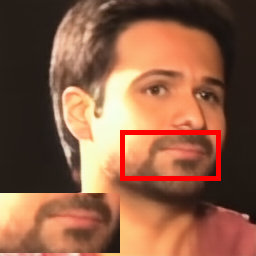} &
    \im{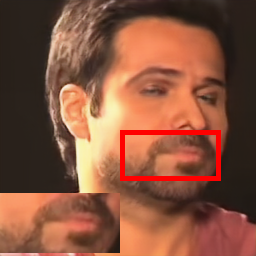} &
    \im{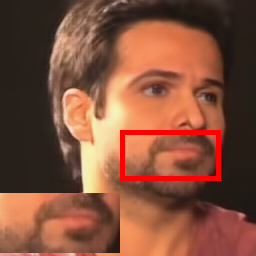} &
    \im{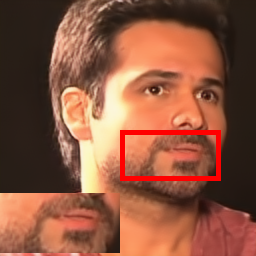} &
    \im{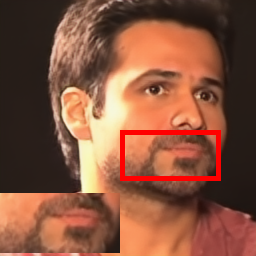}
     \end{tabular}
\caption{Visual illustration of improved reconstruction accuracy with our proposed coding framework. Accumulating references frames and using them to jointly animate the target frames leads to higher accuracy in pixel level details and higher perceptual fidelity.}
\label{fig:visual-comparison}
\end{figure*}

\subsection{Rate-distortion comparison with single-reference GFVC methods}
Our framework that utilizes contrastive learning for multi-reference animation demonstrates superior performance compared to prior methods, both in single reference and multi-reference modes, as evidenced by the reconstruction quality metrics. In Fig.~\ref{fig:rate-distortion}, we show the rate-distortion performance of our framework relative to other GFVC methods and conventional codecs such as HEVC and VVC. Notice that the proposed MRDAC method is used here in an open-loop fashion (i.e., target frames are directly predicted from reference frames), to provide fair comparison with most of previous GFVC approaches. However, our multi-reference approach could be easily integrated into hybrid schemes such as~\cite{konuko2022hdac}, which employ residual coding, to improve prediction performance.
Our proposed MRDAC outperforms other methods showing better perceptual quality, detail preservation, and motion fidelity, as indicated by lower LPIPS, msVGG, and DISTS scores, and higher MS-SSIM, FSIM, and VMAF scores.

\subsection{ Visual Results}

The visual results of our framework shown in Fig.~\ref{fig:visual-comparison}, demonstrate a significant improvement over previous GFVC frameworks. We present three examples to demonstrate the behavior of the proposed frames. In example 1 (top row), we select a target frame where the codec has only a single reference frame. As a result, its higher accuracy in reconstructing the target frame is attributed only to the contrastive learning process. We observe that our framework has a better reconstruction of expressions such as those of the eyes and mouth.Specifically, it has a higher accuracy in aligning the facial features which indicates a higher accuracy in motion prediction and details such as the teeth are better refined compared to the other GFVC frameworks which indicates an improvement in the in-painting mechanism. Note that under single reference animation, the model focuses on realism as shown by the teeth having a good perceptual quality even if they do not match the ground truth. Example 2 (middle row) and example 3 (bottom row) illustrate cases where there are 2 and 3 reference frames, respectively. As the number of reference frames is increased, our method excels in reconstructing the target frames with higher accuracy, preserving details, and maintaining perceptual quality. The visual results confirm the quantitative metrics shown in Fig.~\ref{fig:rate-distortion} in the lower LPIPS, msVGG, and MS-SSIM scores, which indicate better texture and structural preservation. 
\subsection{Bi-direction prediction}
\begin{figure}[htp]
    \centering
\includegraphics[width=0.8\linewidth]{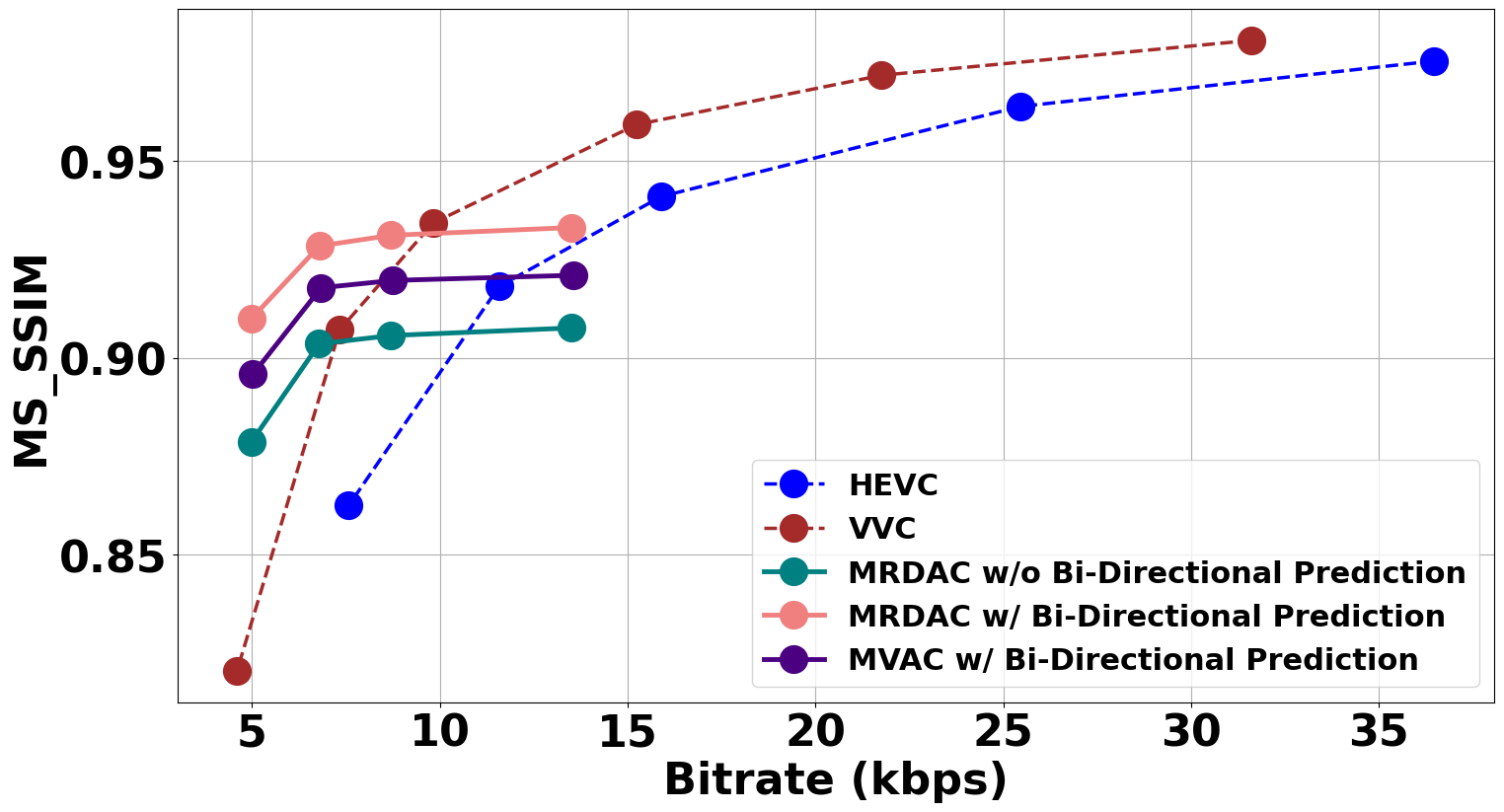}
\caption{Bi-Directional prediction with Multi-Reference Animation}
    \label{fig:bidirectional}
\end{figure}

In Fig.~\ref{fig:bidirectional} we illustrate the relative performance in bi-directional prediction mode, i.e., both past and future frames are used as references. Similar to B frames in conventional codecs, this mode has the potential to increase coding gains, at the cost of augmented structural latency. In this evaluation, we constrain the frameworks to use only a single future reference frame with a maximum delay of 2 seconds. 

From Figure~\ref{fig:bidirectional}, we can make two observations. First, bi-directional prediction can significantly improve visual quality, as intuitively expected. Second, MRDAC has a small but consistent gain over MVAC, likely due to the contrastive loss used for training and weighted aggregation of reference frame features.

\subsection{Reference Frame Selection and Buffering Strategy}
\begin{figure}[htp]
    \centering
\includegraphics[width=0.8\linewidth]{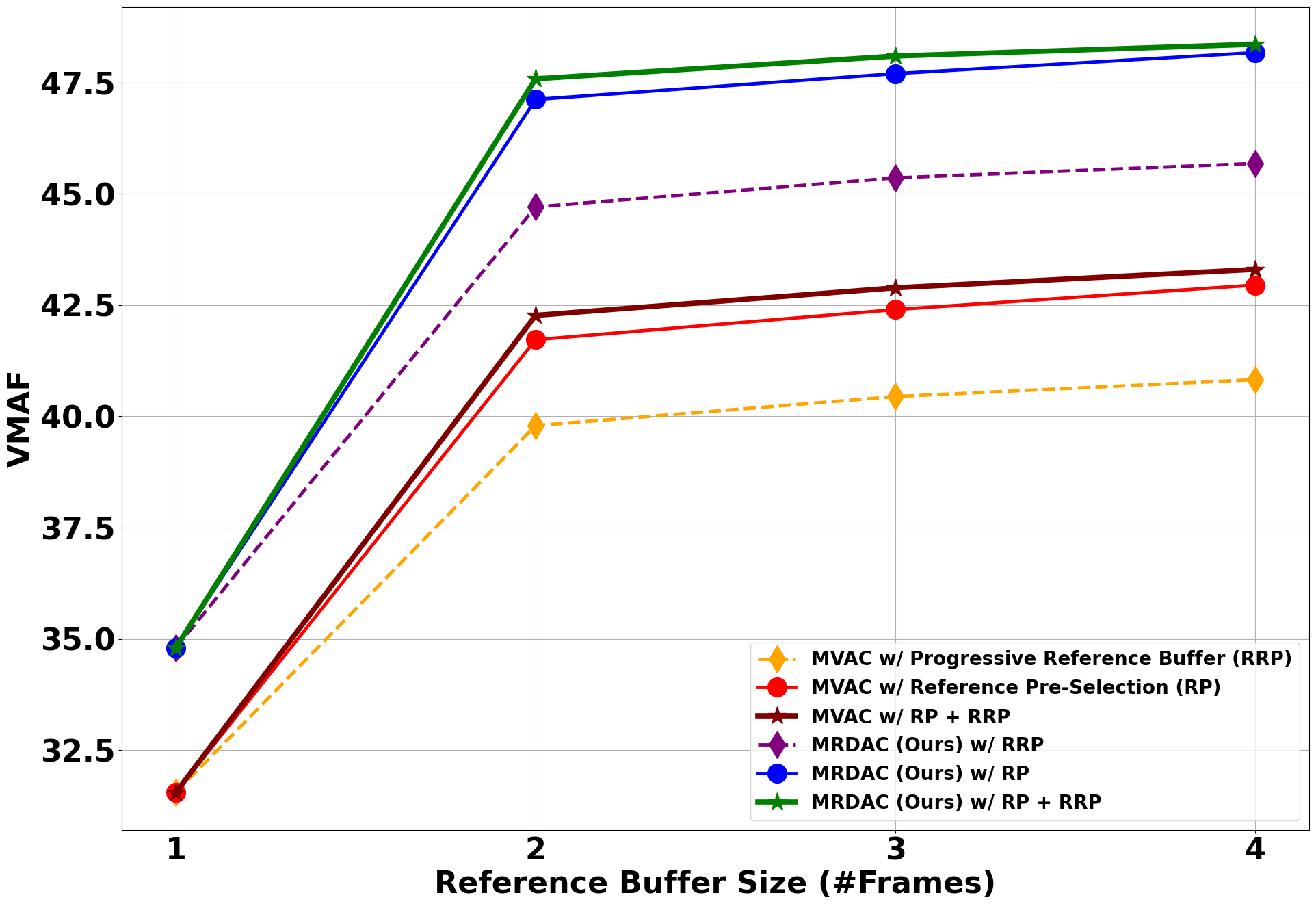}
    \caption{\textbf{Evaluating the effectiveness of coding delay in multi-reference animation:}Using only the past reference frames (RRB) provides incremental gains in but does not effectively capture future poses and motion characteristics. Using a reference pre-selection (RP) provides a better motion model and hence higher accuracy. However using a combination of the two strategies achieves the highest accuracy.}
    \label{fig:pbe-comparison}
\end{figure}

The location and distance of the reference frames relative to each target frame still has significant effects on the reconstruction quality. We study three possible strategies for selecting reference frames: Progressive Reference Buffer (RRB), Reference Pre-Selection (RP), and a combination of both (RP + PRB).
In the RRB strategy, reference frames are accumulated in a buffer as the animation progresses. Previous reference frames are stored and used to animate subsequent target frames. Our results shown in Fig.~\ref{fig:pbe-comparison} indicate that RRB offers incremental gains in reconstruction accuracy as more reference frames are included. 

However, the improvement is limited by the nature of only utilizing past frames, which might not capture future changes in pose or motion effectively.
The RP strategy involves selecting future reference frames before the animation process begins.This pre-selection can be based on various criteria, such as maximizing the diversity of poses, as used in the MVAC~\cite{volokitin2022mvac} framework. For simplicity, we select future reference frames at fixed uniform distances.  This strategy provides a broader context by incorporating future reference frames, allowing the animation model to predict target frames more accurately by anticipating future motions and changes. As discussed in the previous section, RP significantly enhances reconstruction accuracy compared to RRB, at the cost of a coding delay that can be problematic for real-time applications like video conferencing.

Combining the strengths of RRB and RP, the RP+rRB strategy uses both past and future reference frames to animate the target frames. This hybrid approach ensures that most target frames are situated between two or more reference frames, providing a comprehensive context for accurate prediction. We show in Fig.~\ref{fig:pbe-comparison} that this combination not only enhances reconstruction fidelity by leveraging a full temporal spectrum but can also be used to mitigate the coding delay associated with purely future reference selection. By using progressive buffering for past frames and pre-selecting a limited number of future frames, we could achieve a balance that supports high-quality reconstruction while maintaining practical feasibility for real-time applications. In Tab.~\ref{tab:complexity} we provide approximate model complexity for the GFVC methods based on the number of trainable parameters and the number of reference frames. Note that our framework shares a similar architecture to MVAC~\cite{volokitin2022mvac} and thus the performance difference is exclusively attributed to weighted aggregation and contrastive learning.


\begin{table}[ht]
\setlength{\tabcolsep}{2pt}
\small
\begin{center}
\caption{\textbf{ Approximated model complexity as a function of the number of reference frames.}  }
\begin{tabular}{*{3}{c}}
\toprule
&\#Params (M)&{\textit{\bf {\textbf{KMacs/Pixel }} }}\\ 
\midrule
\textbf{CFTE} & 44.9 & 1054 \\
\textbf{HDAC} & 19.6 & 1510 \\
\textbf{DAC} & 7.4 & 615\\
\textbf{\textcolor{teal}{MRDAC}}(\textbf{1 Ref.}) & 9.72 & 762 \\
 \textbf{\textcolor{teal}{MRDAC}}(\textbf{2 Ref.}) & 9.72 & 1146 \\
 \textbf{\textcolor{teal}{MRDAC}}(\textbf{4 Ref.}) & 9.72 & 1834 \\
  \bottomrule
\end{tabular}
\label{tab:complexity}
\end{center}
\vspace{-0.5cm}
\end{table}

\section{Conclusion}
In this paper, we addressed the challenge of reconstructing target frames from video sequences, particularly under conditions of large head motions using multiple reference frames. We proposed a novel method utilizing contrastive learning for multi-reference animation, which improves the robustness and accuracy of frame reconstruction. Our approach enhances feature representation and aggregation through a contrastive loss function, ensuring consistency and fidelity in the generated frames. Quantitative and qualitative results demonstrate that our method outperforms previous GFVC frameworks, as evidenced by superior metrics such as lower LPIPS, msVGG, and DISTS scores, and higher MS-SSIM, FSIM, and VMAF scores. We also study the trade-off between latency and prediction quality, showing that the proposed approach using contrastive learning obtains better performance compared to a similar multi-reference GFVC scheme, MVAC, which does not employ contrastive learning.

\section{Acknowledgments}
This work was funded by LabexDigiCosme - Université Paris-Saclay. This work was performed using HPC resources from GENCI-IDRIS and LTCI-T\'el\'ecom Paris.

\bibliographystyle{IEEEtran}
\bibliography{ref}

\end{document}